\documentclass[10pt,a4papper]{article}
\usepackage{indentfirst}
\usepackage[]{graphicx}
\usepackage{geometry} 
\begin{document}

\title{\textbf{The polarization effects of the process $\tau \rightarrow K^{-} \pi^{0} \nu_{\tau}$ in the Nambu--Jona-Lasinio model}}
\author{A. A. Pivovarov\footnote{tex$\_$k@mail.ru}, O. V. Teryaev\footnote{teryaev@theor.jinr.ru}, \\
\small
\emph{Bogoliubov Laboratory of Theoretical Physics, Joint Institute for Nuclear Research, Dubna, 141980, Russia}}
\maketitle
\small

\begin{abstract}
The polarization effects of the processes $\tau \rightarrow K^{-} \pi^{0} \nu_{\tau}$ are described in the framework of the Nambu--Jona-Lasinio model. The intermediate vector meson $K^{*}(892)$ is taken into account. The contribution of these effects to the differential decay width
is obtained.
\end{abstract}
\large
\section{Introduction}
The research of the hadronic $\tau$-decays is facing difficulties due to inapplicability of the
perturbation theory of quantum chromodynamics at the energy region bellow 2 GeV ($m_{\tau} = 1.777$ GeV).
Thus, one has to use phenomenological models. The most of them are based on the chiral symmetry of strong
interractions and on the vector dominance methods \cite{Finkemeier:1996dh, Li:1996md, Jamin:2006tk, Escribano:2014joa}.
The methods of calculation of angular distributions of the differential widths of the polarized $\tau$-lepton are presented in \cite{Gakh:2015hra}.

In the present paper, we consider the transverse polarization effects in
the framework of the Nambu--Jona-Lasinio (NJL) model.

The NJL model \cite{Volkov:1986zb, Ebert:1985kz, Klevansky:1992qe, Volkov:2006vq} is intended for description of
the four meson nonets in the ground states. Its new version, the extended NJL model
\cite{Volkov:2006vq, Volkov:1996br, Volkov:1996fk, Volkov:1999yi} allow one to describe mesons
in the first radially-excited states. Unlike other phenomenological models, they include a minimal amount of model parameters
and does not require insertion of arbitrary parameters for description of the specific processes.

Recently, several hadronic $\tau$-decays were calculated in the framework of these models, specifically
$\tau \rightarrow \pi \omega \nu_{\tau}$ \cite{Volkov:2012gv}, $\tau \rightarrow (\eta, \eta') 2\pi$ \cite{Volkov:2013zba},
$\tau \rightarrow (\pi, \pi(1300)) \nu_{\tau}$ \cite{Ahmadov:2015zua}, $\tau \rightarrow K^{-} \pi^{0} \nu_{\tau}$ \cite{Volkov:2015vij},
$\tau \rightarrow (\eta,\eta')K^{-}\nu_{\tau}$ \cite{Volkov:2016ebq}, $\tau \rightarrow K^{0}K^{-}\nu_{\tau}$ \cite{Volkov:2016yil}.
However, the polarization effects were not taken into account.

In the present work, the process $\tau \rightarrow K^{-} \pi^{0} \nu_{\tau}$
is calculated in the framework of the NJL model with considering of the polarization of $\tau$-lepton.

\section{The Lagrangian of the NJL model for the mesons $K^{\pm}, \pi^{0}, K^{*\pm}$}
In the NJL model, the quark-meson interaction Lagrangian for pseudoscalar $K^{\pm}, \pi^{0}$ and
vector $K^{*\pm}$ mesons in the ground states takes the form:

\begin{eqnarray}
\Delta L_{int} & = & \bar{q}\left[i\gamma^{5}\sum_{j = \pm}\lambda_{j}g_{K}K^{j}
+ i\gamma^{5}\lambda_{3}g_{\pi}\pi^{0} \right. \nonumber \\
&&\left.+ \frac{1}{2}\gamma^{\mu}\sum_{j = \pm}\lambda_{j}g_{K^{*}}K^{*j}_{\mu} \right]q,
\end{eqnarray}
where $q$ and $\bar{q}$ are the u-, d- and s- constituent quark fields with masses $m_{u} = m_{d} = 280$ MeV,
$m_{s} = 420$ MeV \cite{Volkov:1999yi,Volkov:2001ns}, $K^{\pm}$, $\pi$ and $K^{*\pm}$ are
the pseudoscalar and vector mesons.

The matrices:
\begin{eqnarray}
\lambda_{+} & = &\sqrt{2} \left(\begin{array}{ccc}
0 & 0 & 1\\
0 & 0 & 0\\
0 & 0 & 0
\end{array}\right),\quad
\lambda_{-} = \sqrt{2} \left(\begin{array}{ccc}
0 & 0 & 0\\
0 & 0 & 0\\
1 & 0 & 0
\end{array}\right),\nonumber \\
\lambda_{3} & = & \left(\begin{array}{ccc}
1 & 0  & 0\\
0 & -1 & 0\\
0 & 0  & 0
\end{array}\right).
\end{eqnarray}

The coupling constants:
\begin{eqnarray}
g_{K} & = & \left(\frac{4}{Z_{K}}I_{2}(m_{u},m_{s})\right)^{-1/2} \approx 3.77, \nonumber\\
g_{K^{*}} & = & \left(\frac{2}{3}I_{2}(m_{u},m_{s})\right)^{-1/2} \approx 6.81, \nonumber\\
g_{\pi} & = & \left(\frac{4}{Z_{\pi}}I_{2}(m_{u},m_{u})\right)^{-1/2} \approx 3.02, \nonumber\\
\end{eqnarray}

where
\begin{eqnarray}
Z_{\pi} & = & \left(1 - 6\frac{m^{2}_{u}}{M^{2}_{a_{1}}}\right)^{-1} \approx 1.45, \nonumber\\
Z_{K} & = & \left(1 - \frac{3}{2}\frac{(m_{u} + m_{s})^{2}}{M^{2}_{K_{1}}}\right)^{-1} \approx 1.83,
\end{eqnarray}
$Z_{\pi}$ is the factor corresponding to the $\pi - a_{1}$ transitions,
$Z_{K}$ is the factor corresponding to the $K - K_{1}$ transitions,
$M_{a_{1}} = 1230$ MeV, $M_{K_{1}} = 1272$ MeV \cite{Agashe:2014kda} are the masses
of the axial-vector $a_{1}$ and $K_{1}$ mesons, and the integral $I_{2}$ has the following form:
\begin{eqnarray}
&I_{2}(m_{1}, m_{2}) =
-i\frac{N_{c}}{(2\pi)^{4}}\int\frac{\theta(\Lambda_{4}^{2} - \vec{k}^2)}{(m_{1}^{2} - k^2)(m_{2}^{2} - k^2)}
\mathrm{d}^{4}k,&
\end{eqnarray}
$\Lambda_{4} = 1.25$ GeV is the cut-off parameter \cite{Volkov:1986zb}.

All these parameters were calculated earlier and are standard for NJL model.

\section{The polarization effects of the process $\tau \rightarrow K^{-} \pi^{0} \nu_{\tau}$}
\begin{figure}[h!]
\centerline{\includegraphics[scale = 0.7]{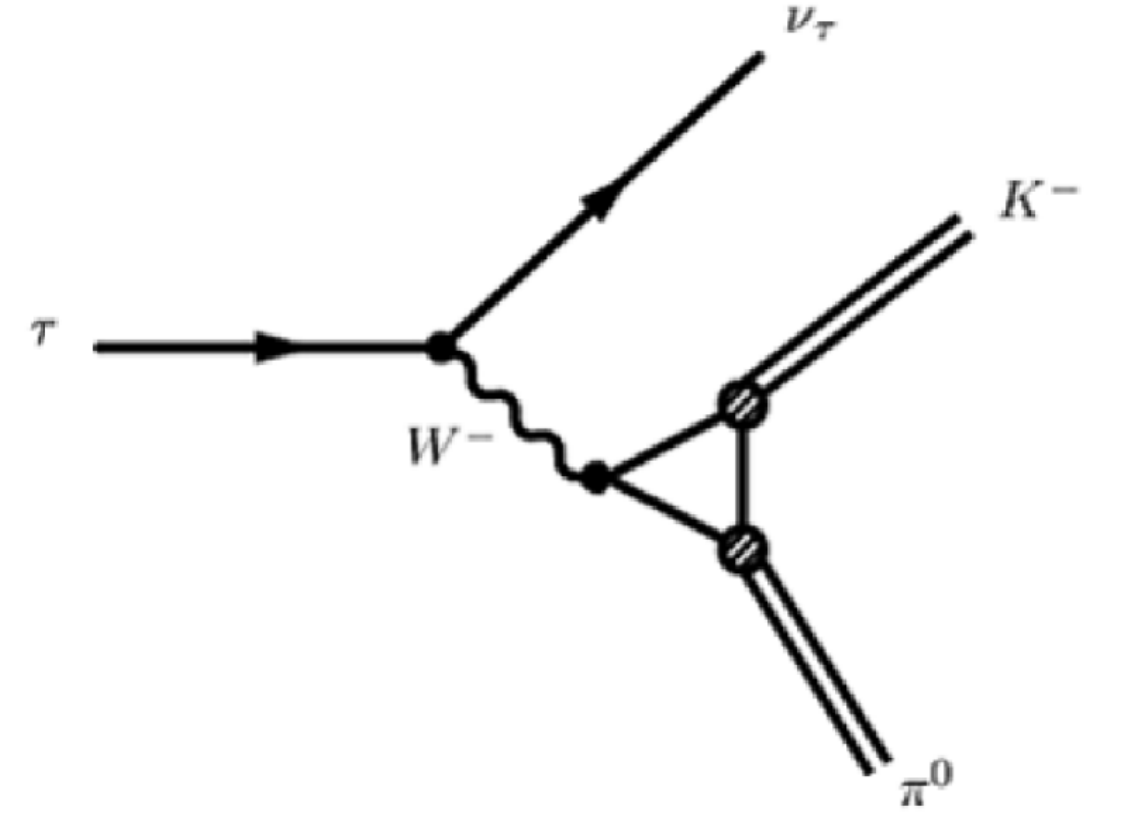}}
\caption{The decay $\tau \rightarrow K^{-}\pi^{0}\nu_{\tau}$ with intermediate $W$-boson (contact diagram).}
\label{Contact}
\end{figure}

\begin{figure}[h!]
\centerline{\includegraphics[scale = 0.7]{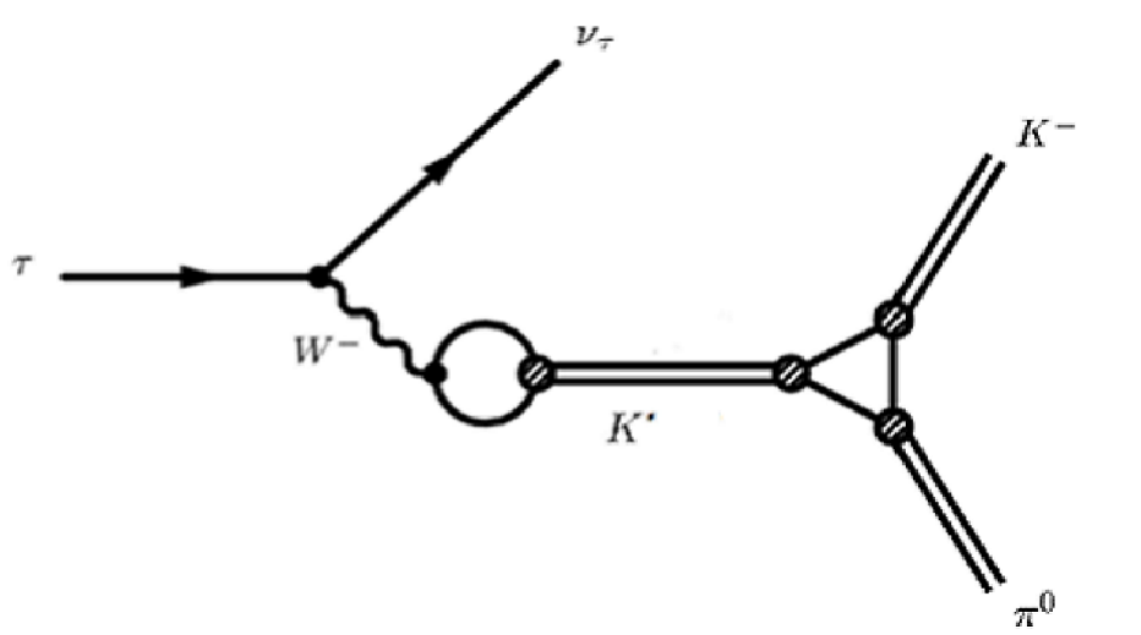}}
\caption{The decay $\tau \rightarrow K^{-}\pi^{0}\nu_{\tau}$ with intermediate vector $K^{*}(892)$ meson.}
\label{Intermediate}
\end{figure}

The diagrams of the decay $\tau \rightarrow K^{-} \pi^{0} \nu_{\tau}$ are shown in Fig.~\ref{Contact} and Fig.~\ref{Intermediate}.


The excited vector meson $K^{*}(1410)$ gives insignificant contribution. Therefore, one can to use the standard NJL model.
The amplitude for this process was obtained in \cite{Volkov:2015vij}:
\begin{eqnarray}
&&T = -\frac{i}{2} G_{F}V_{us} Z_{K} \frac{g_{\pi}}{g_{K}} l^{\mu} \nonumber\\
&&\times \left\{g_{\mu\nu} + \frac{g_{\mu\nu}\left[q^{2} - \frac{3}{2}(m_{s} - m_{u})^{2}\right] - q_{\mu}q_{\nu}}
{M_{K^{*}}^{2} - q^{2} - i\sqrt{q^{2}}\Gamma_{K^{*}}}\right\} (p_{K} - p_{\pi})^{\nu},
\end{eqnarray}
where $G_{F} = 1.16637 \cdot 10^{-11}$ MeV$^{-2}$ is the Fermi constant, $V_{us} = 0.2252$ is the element of the Cabbibo-Kobayashi-Maskawa matrix,
$l^{\mu} = \bar{\nu}_{\tau}\gamma^{\mu}\tau$ is the vector part of the lepton current, $q = p_{K} + p_{\pi}$, $M_{K^{*}} = 896$ MeV,
$\Gamma_{K^{*}} = 46$ MeV are the mass and the full width of the vector meson \cite{Agashe:2014kda}.

For the purpose of considering the polarization of $\tau$-lepton, one has to use the relation:
\begin{eqnarray}
u_{\tau}\bar{u}_{\tau} \rightarrow \frac{1}{2} \left[(p_{\tau}\gamma) + m_{\tau}\right]\left[1 - \gamma^{5}(a\gamma)\right],
\end{eqnarray}
where $a$ is the polarization vector. One can put $|{\bf a}| = 1$.

The taking into account the polarization gives us the additional term containing the antisymmetric tensor
$\epsilon^{p_{\tau}ap_{\pi}p_{K}}$.

After integration over the neutrino momentum, we obtain the differential width as a function of invariant mass of the final
mesons, ratio of the energies of the mesons and the polarization vector of $\tau$-lepton. To estimate the influence of the accounting of
polarization on the differential width we calculate the relation:

\begin{eqnarray}
\label{Relation}
&&\frac{\frac{d\Gamma}{d^{3}p_{\pi}d^{3}p_{K}}(M_{inv}, \varepsilon, a) - \frac{d\Gamma}{d^{3}p_{\pi}d^{3}p_{K}}(M_{inv}, \varepsilon, - a)}
{\frac{d\Gamma}{d^{3}p_{\pi}d^{3}p_{K}}(M_{inv}, \varepsilon, a) + \frac{d\Gamma}{d^{3}p_{\pi}d^{3}p_{K}}(M_{inv}, \varepsilon, - a)} \nonumber\\
&&= -4m_{\tau}M_{inv}\Gamma_{K^{*}}(\varepsilon + 1)^{2}(M_{K}^{2} - M_{\pi}^{2})\epsilon^{p_{\tau}ap_{\pi}p_{K}} \nonumber\\
&&\times \left\{\left(M_{K^{*}}^{2} - 3(m_{s} - m_{u})^{2}/2\right)^{2}\left[4M_{K}^{2}(\varepsilon + 1)(M_{inv}^{2}\varepsilon - m_{\tau}^{2})\right.\right. \nonumber\\
&&+ m_{\tau}^{2}(M_{inv}^{2}(\varepsilon(3\varepsilon - 2) + 3) - 4M_{\pi}^{2}\varepsilon(\varepsilon + 1)) \nonumber\\
&&\left.+4M_{inv}^{2}(M_{\pi}^{2}(\varepsilon + 1) - M_{inv}^{2}\varepsilon)+ m_{\tau}^{4}(\varepsilon - 1)^{2}\right] \nonumber\\
&&+ 2\left(M_{K^{*}}^{2} - 3(m_{s} - m_{u})^{2}/2\right)m_{\tau}^{2}(\varepsilon + 1)(M_{K}^{2} - M_{\pi}^{2}) \nonumber\\
&&\times \left[2(M_{K}^{2} - M_{\pi}^{2})(\varepsilon + 1) + (\varepsilon - 1)(m_{\tau}^{2} + M_{inv}^{2})\right] \nonumber\\
&&+ M_{K}^{2}m_{\tau}^{2}(\varepsilon + 1)^{2}(m_{\tau}^{2} - M_{inv}^{2})(M_{K}^{2} - 2M_{\pi}^{2}) \nonumber\\
&&- 4M_{K}^{2}(\varepsilon + 1) M_{inv}^{2}\Gamma_{K^{*}}^{2}(m_{\tau}^{2} - M_{inv}^{2}\varepsilon) \nonumber\\
&&+ M_{\pi}^{4}m_{\tau}^{2}(m_{\tau}^{2} - M_{inv}^{2})(\varepsilon^{2} + 1) + 4M_{\pi}^{2}M_{inv}^{4}\Gamma_{K^{*}}^{2} \nonumber\\
&&+ m_{\tau}^{2}\varepsilon^{2}M_{inv}^{2}\Gamma_{K^{*}}^{2}(-4M_{\pi}^{2} + m_{\tau}^{2} + 3M_{inv}^{2}) \nonumber\\
&&+ 2\varepsilon\left[M_{\pi}^{4}m_{\tau}^{2}(m_{\tau}^{2} - M_{inv}^{2}) - M_{inv}^{2}\Gamma_{K^{*}}^{2}(2M_{\pi}^{2}(m_{\tau}^{2} - M_{inv}^{2})\right. \nonumber\\
&&\left.\left. + m_{\tau}^{4} + m_{\tau}^{2}M_{inv}^{2} + 2M_{inv}^{4})\right] + m_{\tau}^{2}M_{inv}^{2}\Gamma_{K^{*}}^{2}(m_{\tau}^{2} + 3M_{inv}^{2})\right\}^{-1},
\end{eqnarray}
where $M_{inv}$ is the invariant mass and $\varepsilon = \frac{\varepsilon_{\pi}}{\varepsilon_{K}}$ is the relation of the energies
of the final mesons.

The appearance of kaon width in the numerator is due to the fact, that asymmetry is proportional to
the phase shift between invariant amplitudes (similar effects in QCD were considered in \cite{Efremov:1981sh}),

In the rest system of $\tau$-lepton, the time component of the vector $a$ disappears and the expression for the antisymmetric tensor takes the form:
\begin{eqnarray}
\epsilon^{p_{\tau}ap_{\pi}p_{K}} = m_{\tau}|{\bf a}||{\bf p}_{\pi}||{\bf p}_{K}| \textrm{sin}\alpha \textrm{cos}\beta,
\end{eqnarray}
where
$$\alpha  = \textrm{Arccos} \frac{M_K^2+M_\pi^2+2 \varepsilon_K \varepsilon_\pi  -M_{inv}^2}{2 |{\bf p}_{\pi}||{\bf p}_{K}|} $$
is the angle between the momenta of the $\pi$- and K-mesons,
$\beta$ is the angle between the
polarization vector and the normal to the plane defined by these momenta.

\begin{figure}[h!]
\centerline{\includegraphics[scale = 0.8]{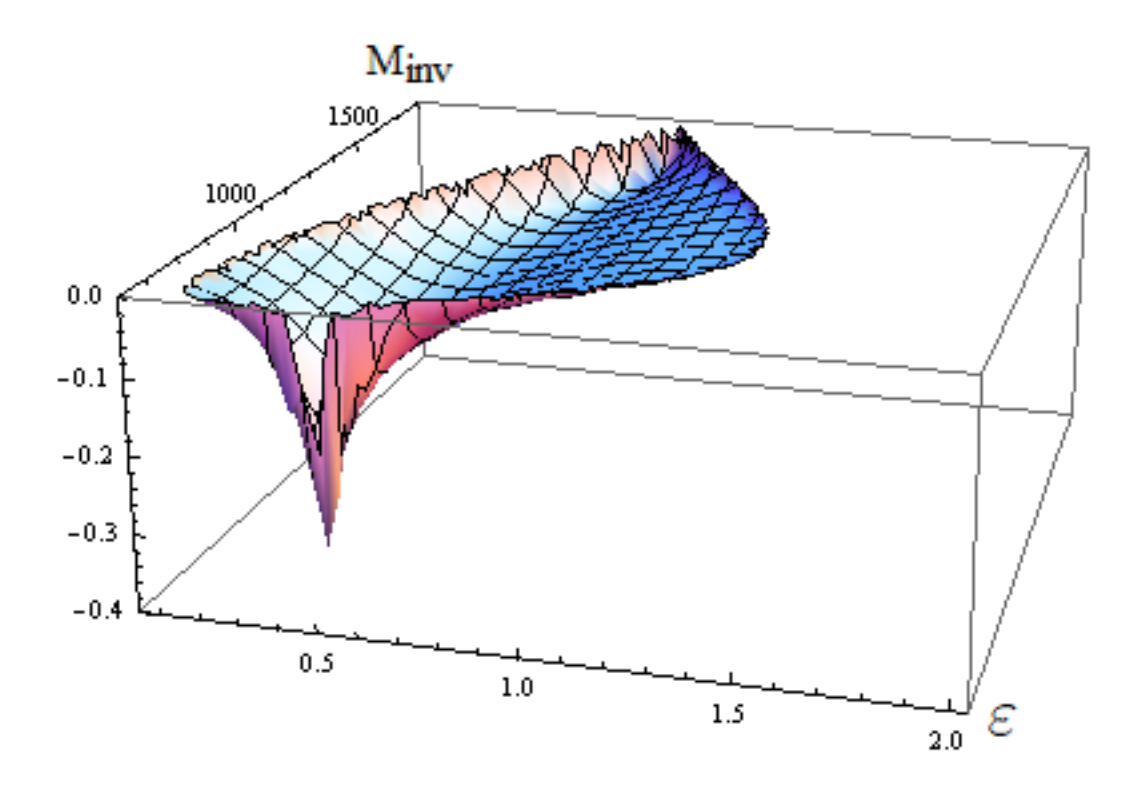}}
\caption{The asymmetry of the differential width due to transverse polarization of $\tau$-lepton}
\label{Diff}
\end{figure}

The dependence of the relation (\ref{Relation}) on $M_{inv}$ and $\varepsilon$ for the case
$\beta = 0^{\circ}$ (transverse polarization providing the maximal asymmetry) is shown in Fig.~\ref{Diff}.

\section{Conclusion}
In the present work, the polarization effects of the decay $\tau \rightarrow K^{-} \pi^{0} \nu_{\tau}$ in the framework of the Nambu--Jona-Lasinio model were considered. The contribution of these effects to the differential width is obtained. The resulting asymmetry for transverse polarized lepton is of order $10 \%$ and can be measured in high statistics experiments. It provides the new sensitive test for
NJL model.

\section*{Acknowledgments}
We are grateful to M. K. Volkov for useful discussions; this work is supported by the RFBR grant, 14-01-00647.

\end{document}